\input amstex
\input xy
\input epsf
\xyoption{all}
\documentstyle{amsppt}
\document
\magnification=1200
\NoBlackBoxes
\nologo
\hoffset1.5cm
\voffset2cm

\def\Z{\bold{Z}}

\def\r{\roman}

\pageheight {16cm}


\bigskip

\centerline{\bf  ZIPF'S LAW }
\smallskip
\centerline{\bf AND L.~LEVIN'S PROBABILITY DISTRIBUTIONS} 

\medskip
\centerline{\bf Yuri I. Manin}

\medskip

\centerline{\it Max--Planck--Institut f\"ur Mathematik, Bonn, Germany}

\bigskip

{\it ABSTRACT.} Zipf's law in its basic incarnation is an empirical probability distribution 
governing the frequency of usage of words in a language.  As Terence Tao recently remarked,
it still lacks a convincing and satisfactory mathematical explanation. 

In this paper I suggest that at least in certain situations, Zipf's law can be explained as
a special case of   the {\it a priori} distribution introduced and studied by L.~Levin. The Zipf ranking
corresponding to diminishing probability appears then as the ordering determined by the
growing Kolmogorov complexity.

One  argument  justifying this assertion is the appeal to a recent interpretation by Yu.~Manin and M.~Marcolli of
asymptotic bounds for error--correcting codes in terms of phase transition.
In the respective  partition function,  Kolmogorov
complexity of a code plays the role of its energy.

\bigskip

\centerline{\bf 0. Introduction and summary}

\medskip

{\bf 0.1.~Zipf's law.} Zipf's law was discovered as  an empirical observation ([Zi1], [Zi2]): if all words $w_k$  of a language
are ranked according to decreasing frequency of their appearance 
in a representative corpus of texts, then the frequency $p_k$ of $w_k$ is (approximately) inversely proportional
to  its rank $k$: see e.~g.~ Fig.~1 in [Ma1] based upon a corpus containing $4\cdot 10^7$ Russian words.

\smallskip
For various other incarnations of this power exponent $-1$ law in many
different statistical data, cf.~[MurSo] and references therein.

\smallskip

Theoretical models of Zipf's distribution also abound. In the founding texts
of Zipf himself [Zi2], {Zi1], it was suggested that his distribution ``minimizes effort''. Mandelbrot 
in [Mand]  described a concise mathematical framework for producing a model of  Zipf's 
law. Namely, if we postulate and denote by $C_k$ a certain ``cost'' (of producing, using etc.) of the
word of rank $k$, then the frequency distribution $p_k \sim 2^{-h^{-1}C_k}$ minimizes
the ratio $h=C/H$, where $C:=\sum_k p_kC_k$ is the average cost per word,  and $H:= -\sum_kp_k \roman{log}_2 p_k$
is the average entropy: see [Ma2].

\smallskip

We get from this a power law, if  $C_k\sim \roman{log}\,k$.  An additional problem,
what is so special about power $-1$, must be addressed separately. For one possibility,
see [MurSo], sec.~III. In this work, we suggest a different mathematical background
(see the next subsection).

\smallskip

In all such discussions, it is more or less implicitly assumed that empirically observed distributions
 concern  fragments of a potential countable infinity of objects. I also postulate this,
 and work in such a ``constructive world'': see sec. 1.1 below for a precise definition.

\medskip

{\bf 0.2.~Zipf's law from complexity.} In this note we suggest that (at least in some situations) Zipf's law emerges as  the combined 
effect of two factors:

\smallskip

{\it (A) Rank ordering coincides with the ordering with respect to the growing (exponential) Kolmogorov complexity $K(w)$ 
up to a factor  $exp\,(O(1))$.}

\smallskip

More precisely, to define $K(x)$ for a natural number $x\in \Z^+$, we choose a {\it Kolmogorov optimal encoding,} which is a
partial recursive function $u:\, \Z^+ \to \Z^+$, and put $K(x)=K_u(x) := \roman{min}\,\{y\,|\,u(y)=x$.
Another choice of $u$ changes $K_u(.)$ by a factor $exp\,(O(1))$. 

\smallskip

Furthermore, $K(w)$ for elements $w$ of a constructive world is defined as complexity of its number
in a fixed structural numbering (cf. 1.1 below). Changing the numbering, we again change complexity
by a  $exp\,(O(1))$--factor.  

\medskip

{\it (B) The probability distribution producing Zipf's law (with exponent $-1$) is
(an approximation to) the L.~Levin maximal computable from below
distribution: see [ZvLe], [Le] and [LiVi].}

\medskip
If we accept $(A)$ and $(B)$, then Zipf's law follows from two basic properties of Kolmogorov complexity:
\medskip
{\it (a) rank of $w$ defined according to (A) is $exp\,(O(1))\cdot K(w)$.}
\medskip

{\it (b) Levin's distribution  assigns to an object $w$  probability
$\sim KP(w)^{-1}$ where $KP$ is the exponentiated prefix Kolmogorov complexity,
and we have, up to  $exp\, (O(1))$--factors,
$$
K(w)\preceq KP(w)\preceq  K(w)\cdot \roman{log}^{1+\varepsilon} \,K(w)$$}
with arbitrary $\varepsilon >0$.
\smallskip

Slight discrepancy between the growth orders of  $K$ and $KP$ is the reason why
a probability distribution on infinity of objects cannot be constructed from $K$:
the series $\sum_m K(m)^{-1}$ diverges. However, on finite sets of data 
this small discrepancy is additionally masked by the dependence
of both $K$ and $KP$ on the choice of  an optimal encoding. Therefore, when speaking about Zipf's Law, we will mostly disregard this
difference. See also the discussion of the partition function for codes in 0.3 below.

\medskip

{\bf 0.3. Complexity as effort.}  The picture described above agrees with Zipf's motto
``minimization of effort'', but reinterprets the notion of effort: its role is now played
by the logarithm of the Kolmogorov complexity that is by the length of the maximally
compressed description of  an object. This length is not computable, 
but it is the infimum of a sequence of computable functions. 
\smallskip

Such a picture makes sense especially if the objects satisfying
Zipf's distribution, are {\it generated} rather than simply {\it observed.}

\smallskip
Intuitively, whenever
an individual mind, or a society, finds a compressed description of something,
this something becomes usable, and is used more often than other "something"
whose description length is longer. For an expanded version of this metaphor
applied to the history of science, see [Man3].

\smallskip

For words in the initial Zipf's observation, this principle
refers  to ways in which mind/brain generates and uses language.

\medskip

{\bf 0.4.~Relation to previous works.} I am aware of two works where complexity is invoked in relation to Zipf's law:
[Ve] and [MurSo].\footnotemark1
\footnotetext{By the time this text was essentially written, one more article [Del] appeared that suggests
essentially the same relation between Zipf and complexity as this paper.
Prof. Jean--Paul Delahaye kindly drew my attention to it after my article was posted in arXiv.  }

\smallskip

(a)  Briefly, viewpoint of [MurSo] is close to ours, but, roughly speaking, the authors focus on the majority
of objects consisting of  ``almost random'' ones: those whose size is comparable with  Kolmogorov complexity,
and which therefore cannot be compressed. This is justified by
the authors' assumption that the data corpus satisfying Zipf's Law comes from
a sequence of  successive observations over a certain system with stochastic behaviour.

\smallskip
To the contrary, our ranking puts in the foreground those objects whose size might be very large
in comparison with their complexity, because we imagine systems that are {\it generated} 
rather than simply {\it observed}, in the same sense as texts written in various languages are generated
by human brains.

\smallskip

To see the crucial difference between the two approaches on a well understood
mathematical example, one can compare them on the background of error--correcting codes,
following [ManMar].
Each such code $C$ (over a fixed alphabet) determines a point in the unit square of the plane 
{\it (transmission rate, minimal relative Hamming's distance).} The closure of all limit code points
is a domain lying below a certain continuous curve which is called {\it asymptotic bound.}

\smallskip
 If one produces
codes in the  order of growing size, most  code points will form a cloud
densely approximating the so called Varshamov--Gilbert bound that lies strictly below
the asymptotic bound.

\smallskip
To the contrary,  if one produces codes in the order of their Kolmogorov
complexity, their code points will well approximate the picture of the whole domain
under the asymptotic bound: see details in [ManMar]. Moreover, Levin's distribution very naturally leads to
a thermodynamic partition function on the set of codes, and to the interpretation of asymptotic bound as
a phase transition curve: this partition function has the form $\sum_C K(C)^{-s(C)}$ where $s(C)$
is a certain function defined on codes and including as parameters analogs of temperature and density. 
Here one may replace $K$ with $KP$, and freely choose
the optimal family defining complexity: this will have {\it no influence at all} on the form of the phase curve/asymptotic bound.
\smallskip

In this partition function, $\roman{log}\, K(w)$ that is the bit--size of a maximally compressed description of $w$,
plays precisely the role of energy in this partition function, thus validating our suggestion to identify it with
``effort''.

\smallskip

It is interesting to observe that the mathematical problem of generating good error--correcting codes
historically made a great progress in the 1980's with the discovery of algebraic geometric
Goppa codes, that is precisely with the discovery of greatly compressed descriptions of
large combinatorial objects.

\smallskip

To summarize, the class of a priori probability distributions that we are considering here
is {\it qualitatively distinct} from those that form now  a common stock of sociological and sometimes scientific analysis:
cf.~a beautiful synopsis of the latter by Terence Tao in [Ta] who also stresses that
``mathematicians do not have a fully satisfactory and convincing explanation
for how the [Zipf] law comes about and why it is universal''.

\smallskip

(b) We turn now to the paper [Ve], in which  T.~Veldhuizen  considers Zipf's law in an unusual context that did not exist in the days 
when Kolmogorov, Solomonov and Chaitin made their discoveries, but which
provides, in a sense, landscape for an industrial incarnation of  complexity.
Namely, he studies actual software and software libraries and 
analyzes  possible profits from software reuse. Metaphorically, this is a picture of human culture
whose everyday existence depends on a continuous reuse of treasures
created by researchers, poets, philosophers.
\smallskip

Mathematically, reuse furnishes new tools of
compression:  roughly speaking, a function $f$ may have a very large
Kolmogorov complexity, but the length of the library address of its program
may be short, and only the latter counts 
if one can simply copy the program from the library. 

\smallskip

In order to create a mathematical model of reuse and its Zipf's landscape along the lines of this
note, I need to define the mathematical notion
of {\it relative Kolmogorov complexity $K(f|F)$.} This notion goes back to Kolmogorov
himself and  is well known in the case when 
$f, F$ are finite combinatorial objects such as strings or integers (cf.~[LiVi]). 

\smallskip

In the body of the paper,
I generalize this definition to the case of  {\it a library $F$} of programs.
The library may even contain uncomputable oracular data, and thus we include into the complexity landscape
 oracle--assisted computations.

\medskip

{\bf 0.5.~Some justifications.} Consider some experimental data demonstrating
the dependance of Zipf's rank from complexity in the most natural environment:
when we study the statistics not of all words, but only {\it numerals}, the names of numbers.

\smallskip

Then in our model we expect that:

\smallskip

(i) Most of the numbers $n$, those that are Kolmogorov "maximally complex", will appear with
probability comparable with $n^{-1}\,({\roman{log}\,n})^{-1-\varepsilon}$, with a small $\varepsilon$:
``most large numbers appear with frequency inverse to their size'' (in fact, somewhat smaller one).

\smallskip

(ii) However, frequencies of  those numbers that are Kolmogorov very simple, such as $10^3$ (thousand),
$10^6$ (million), $10^9$ (billion), must produce sharp local peaks in the graph of $(p_n)$.

\smallskip

The reader may compare these properties of the  discussed class of Levin's  distributons, which can be
called {\it a priori distributions}, with the observed frequencies of numerals 
in printed and oral texts in several languages, summarized in Dehaene, [De],  p.~111,  Figure 4.4.
(Those parts of the Dehaene and Mehler graphs  in the book [De] that refer
to large numbers, are somewhat misleading: they might create an impression that frequencies of the numerals, say,
between $10^6$ and $10^9$ smoothly interpolate between those of   $10^6$ and $10^9$ themselves,
whereas in fact they abruptly drop down. See, however, a much more detailed discussion in [DeMe].)
The article [Del] also quotes ample empirical data obtained by Google search. 
\smallskip
To me, the observable {\it qualitative} agreement between suggested theory and observation looks  convincing: brains and their societies
do follow predictions of a priori probabilities. Of course, one has to remember that
compression degrees that can be achieved by brains and civilisations
might produce {\it quantitatively} different distributions at the initial segments of a Kolmogorov
Universe, because here dependence of the complexity of objects  on the complexity of generating them
mechanisms (``the culture code'')  becomes pronounced. 
\smallskip

There is no doubt that many instances of empiric Zipf's laws will not be reducible
to our complexity source. Such a reduction of the Zipf law for all words
might require for its justification some  neurobiological data: cf. [Ma1], appendix A
in the arXiv version.

\smallskip

Another interesting possible source of Zipf's law was considered in a recent paper [FrChSh].
The authors suggested that Zipf's rank of an object, member of a certain universe,
might coincide with its PageRank with respect to  an appropriate directed network
connecting all objects. This mechanism generally produces a power law, but not necessarily 
exactly Zipf's one.

\smallskip
In any case,  the appeal to the uncomputable degree of maximal compression in our model of Zipf--Levin distribution is
exactly what can make such a model an eye--opener. 

\medskip

{\bf 0.6. Fractal landscape of the Kolmogorov complexity and universality of Zipf's law.} A graph of logarithmic
Kolmogorov complexity of integers $k$ (and its prefix versions) looks as
follows: most of the time it follows closely  the graph of $\roman{log}\,k$,
but infinitely often it drops down, lower than any given computable function:
see [LiVi], pp. 103, 105, 178. The visible``continuity" of this graph
reflects the fact that complexity of $k+1$ in any reasonable encoding is almost the same
as complexity of $k$.

\smallskip

However, such a picture cannot convey  extremely rich
self--similarity properties of complexity. The basic
fractal property is this: if one takes any infinite decidable subset
of $\Z^+$ in increasing order and restricts the complexity graph on this subset,
one will get the same complexity relief as for the whole $\Z^+$:
in fact, for any recursive bijection $f$ of $\Z^+$ with a subset of $\Z^+$
we have $K(f(x)) = exp (O(1))\cdot K(x)$.

\smallskip

Seemingly, this source of ``fractalization'' might have a much wider influence: see
[NaWe] and related works.

\smallskip

If we pass from complexity to a Levin's distribution, that is, basically, invert the 
values of complexity, these fractal properties survive. 

\smallskip

This property might be accountable for "universality"
of Zipf's law, because it can be read as its extreme stability
with respect to the passage to  various sub--universes
of objects, computable renumbering of objects etc.
It is precisely this stability that underlies the suggestion made
in [HuYeYaHu]: to use deviations from Zipf's Law in patterns
detected in vast databases in order to identify potential fraud
records by auditors.

\smallskip

In the same way, the picture of random noise
in a stable background is held responsible for universality
of normal distribution.

\medskip

{\bf 0.7. Plan of the paper.}  In the main body of the paper, I do not argue anymore 
that complexity may be a source of Zipf's distribution. Instead, I sketch
mathematics of complexity in a wider context, in order to make it applicable
in the situations described in [Ve].
\smallskip
In sec.~1, I define the notion of
Kolmogorov complexity relative to an {\it admissible family} of (partial) functions.
The postulated properties of such admissible families should reflect
our intuitive idea about library reuse and/or oracle--assisted computations. 
\smallskip
In sec.~2, I suggest a formalisation of the notion of computations that  (potentially) produce
admissible families. It turns out that categorical and operadic notions 
are useful here, as it was suggested in [Man1], Ch.~IX.
\medskip

{\it Acknowledgements.} I first learned about Zipf's Law from (an early version of) the paper [Ma] by D.~Yu.~Manin,
and understood better the scope of  my constructions trying to answer his questions.

\smallskip
The possibility that Zipf's law reflects  Levin's distribution occurred to me
after looking at the graphs in the book [De] by S.~Dehaene. Professor Dehaene 
also kindly sent me the original paper [DeMe].  C.~Calude read several versions of this
article and stimulated  a better presentation of my arguments. A.~D\"oring corrected several
misprints in an earlier version.

\smallskip
An operadic description (cf.  [BoMan]) of
a set of programs for computation of
(primitive) recursive functions was discussed in  [Ya], and my old e--mail correspondence with N.~Yanofsky 
helped me to clarify my approach to the formalisation of the notions of reuse and oracles.
Finally, L.~Levin suggested several useful revisions.
I am very grateful to all of them.

\bigskip

\centerline{\bf  1.~Admissible sets of partial functions and relative complexity}

\medskip

{\bf 1.1. Notations and conventions.} We recall here some basic conventions of [Man1], Ch.~V and IX.
Let $X$, $Y$ be two sets. A {\it partial function}  from $X$ to $Y$ is a pair $(D(f),f)$ where $D(f)\subset X$, $f:\,D(f)\to Y$.
We call $D(f)$ {\it the domain} of $f$, and often write simply $f:\,X\to Y$. If $D(f)=\emptyset$, $f$ is called an empty function.
If $D(f)=X$, we sometimes call $f$ a {\it total} function.
If $X$ is one--element set, then non--empty functions $X\to Y$ are canonically identified with elements of $Y$.
Partial functions can be composed in an obvious way: $D(g\circ f ):= f^{-1}(D(g)\cap Im\,(f))$.
Thus we may consider a category consisting
of (some) sets and partial maps between them. 

\smallskip

Put $\bold{Z}^+:=$ the set of positive integers. Then $(\Z^{+})^m$ for $m\ge 1$ can be identified with
the set of vectors $(x_1,\dots ,x_m)$, $x_i\in \Z^+$. By definition,  $(\Z^+)^0$ is an one-element set, say, $\{*\}$.
Any partial function $f:\, (\Z^{+})^m \to  (\Z^{+})^n$ will be called an $(m,n)$--function.
For $m=0$, such a non--empty  function may and will be identified with a vector from  $(\Z^{+})^n$.

\smallskip

Let $X$ be an infinite set. A structure of {\it constructive world} on $X$ is given by a set of bijections $\Cal{N}_X$ called
{\it structure numberings}, $X\to \Z^+$, such that any two bijections in it are related by
a (total) recursive permutation of  $\Z^+$,  and conversely, any composition of a structural numbering
with a recursive permutation is again a structure numbering. Explicitly given finite sets are also
considered  as constructive worlds. (A logically minded reader
may imagine all our basic constructions taking place at the ground floor of
the von Neumann Universe).

\smallskip

Intuitively, $X$ can be imagined as consisting of certain finite Bourbaki structures that can be unambiguously described
and encoded by finite strings in a finite alphabet that form a decidable set of strings,
and therefore also admit a natural numbering. Any two such natural numberings, of course,
must be connected by a computable bijection.

\smallskip

Morphisms between two constructive worlds, by definition, consist of those set--theoretical maps which,
after a choice of structural numberings, become partially recursive functions. Thus, constructive worlds are objects of a category, Constructive Universe.

\smallskip

In order to introduce a formalization of oracle--assisted computations, we
will have to extend the sets of morphisms allowing partial maps
that might be non--computable.
\medskip

{\bf 1.2. Admissible sets of functions.} Consider a  set $\Phi$  of partial functions   $f:\, (\Z^{+})^m \to  (\Z^{+})^n$, $m,n\ge 0$.
We will call $\Phi$ {\it an admissible set}, if it is countable and satisfies the following conditions. 

\smallskip
{\it (i) $\Phi$   is closed under composition and contains all projections (forget some coordinates),
and embeddings (permute and/or add some constant coordinates).}

\smallskip

 Any $(m+1,n)$--function  can be considered as a family of $(m,n)$--functions $(u_k)$:
$u_k(x_1,\dots ,x_m):= u(x_1,\dots ,x_m, k)$.  From (i) it follows that for any $u\in \Phi$ and $k\in \Z^+$, also
$u_k\in \Phi$. Similarly, if $u(x_1,\dots ,x_m)$ is in $\Phi$, then 
$$
U(x_1,\dots,x_m,x_{m+1},\dots ,x_{m+n})\equiv
u(x_1,\dots ,x_m)
$$ 
is in $\Phi$.

\smallskip
{\it (ii) For any $(m,n)$, there exists 
an $(m+1,n)$--function $u\in \Phi$ such that the  family of functions   $u_k:\,(\Z^{+})^m \to  (\Z^{+})^n$, 
contains all $(m,n)$--functions   belonging to $\Phi$.
\smallskip
We will say that such a function $u$ (or family $(u_k)$) is ample.}

\smallskip

{\it (iii) Let $f$ be a total recursive function $f$ whose image is
 decidable, and $f$ defines a bijection between $D(f)$ and  image of $f$.
 Then $\Phi$ contains both $f$ and $f^{-1}$.}

\smallskip

From now on, $\Phi$ will always denote an admissible family.

\medskip

{\bf 1.3. Complexity relative to a family.}  Choose an $(m+1,n)$--function $u\in \Phi$ 
and consider it as a family of $(m,n)$--functions $u_k$ as above.
For any $(m,n)$--function $f\in \Phi$,  put $K_u^{\Phi}(f) = \roman{min}\,\{k\,|\, f=u_k\}.$
The r.h.s. is interpreted as $\infty$ if there is no such $k$.
We will call such a family $u$ {\it Kolmogorov optimal in $\Phi$}, if for any other $(m+1,n)$--function
$v$ there is a constant $c_{u,v}$ such that for all $(m,n)$--functions $f\in \Phi$ we
have $K_u^{\Phi}(f)\le c_{u,v}K_v^{\Phi}(f)$.

\medskip

{\bf 1.4. Theorem.} {\it a) If $\Phi$ contains an ample family of $(m+1,n)$--functions, than it contains also
a Kolmogorov optimal family of $(m,n)$--functions.

\smallskip
b) If $u$ and $v$ are two  Kolmogorov optimal families of $(m,n)$--functions, then 
$$
c_{v,u}^{-1}\le K_u^{\Phi}(f)/K_v^{\Phi}(f)\le c_{u,v}.
$$
}

{\bf Proof.}  Let $\theta :\, \Z^+\times \Z^+\to \Z^+$ be a total recursive bijection between $\Z^+\times \Z^+$
and a decidable subset of $\Z^+$. Assume moreover that $\theta (k,j)\le k\cdot\phi (j)$ for some
$\phi :\Z^+\to\Z^+$. Choose any ample family $U\in \Phi$ of $(m+1,n)$--functions and put
$$
u(x_1,\dots ,x_m,k):=U(x_1,\dots ,x_m,\theta^{-1}(k)).
$$
Then $u$ is ample and optimal, with the following bound for the constant $c_{u,v}$:
$$
c_{u,v}\le \phi (K^{\Phi}_U(v)).
\eqno(1.1)
$$
In fact, it suffices to consider such $v$ that $f$ occurs in $(v_k)$. Then
$$
f(x_1,\dots ,x_m)=v(x_1,\dots ,x_m, K_v^{\Phi}(f))
$$
$$
=U(x_1,\dots ,x_m, K_v^{\Phi}(f), K_U^{\Phi}(v))
$$
so that
$$
K_u^{\Phi}(f) \le  \theta (K_v^{\Phi}(f) ,  K_U^{\Phi}(v) ) \le   K_v^{\Phi}(f) \phi( K_U^{\Phi}(v)).
$$

\medskip

{\bf 1.5. Constants related to Kolmogorov complexity estimates.}  In the inequality (1.1)  
estimating the dependance of Kolmogorov complexity on the choice of encoding, two factors play
the central roles.  
\smallskip
One is $K_U^{\Phi}(v)$. Its effective calculation depends on the possibility
of translating a program for $v$ into a program given by $U$. In the situation where
$\Phi$ consists of all partial recursive functions,  such a compilation can be performed 
if $U$ satisfies a property that is stronger  than ampleness: cf. [Ro] and [Sch]
where such families are discussed and constructed.
\smallskip

Another factor is the growth rate of $\phi$.  Below we will show how the task of
optimization of $\phi$ can be seen in the context of Levin's
distributions, reproducing an argument from [Man2].
\medskip

{\bf 1.5.1. Slowly growing numberings of $(\Z^+)^2.$} Let $R=(R_k\,|\,k\in\Z^+)$ 
be a sequence of positive  numbers tending to infinity
with $k$. For $M\in \Z^+$, put
$$
V_R(M):=\{(k,l)\in (\Z^+)^2 \,|\, kR_l\le M\}.
$$
Clearly,
$$
\r{card}\,V_R(M)\le \sum_{l=1}^{\infty} \left[\frac{M}{R_l}\right] < \infty\,,
$$
where $[a]$ denotes the integral part of $a$.

\smallskip

We have 
$$
V_R(M)\subset V_R(M+1),\  (\Z^+)^2=\cup_{M=1}^{\infty} V_R(M).
$$

Therefore we can define a  bijection $N_R:\,(\Z^+)^2 \to \Z^+ $ in the following
way: $N_R(k,l)$ will be the rank of $(k,l)$ in the total ordering
$<_{R}$ of $(\Z)^2$ determined inductively by the following rule:
$(i,j)<_R(k,l)$ iff one of the following alternatives holds: 
\smallskip
(a) $iR_j<kR_l$;
\smallskip
(b) $iR_j=kR_l$ and $j<l$;
\medskip

{\bf 1.5.2. Proposition.} {\it The numbering $N_R$ is well defined
and has the following property: all elements of $V_R(M+1)\setminus 
V_R(M)$ have strictly larger ranks than those of $V_R(M)$. Moreover:
\smallskip

(i)  If the set $\{\,(q,l)\in \bold{Q}\times\bold{Z}^+\,|\,q\ge R_l\}$
is enumerable (image of a partial recursive function), then
$N_R$ is computable (total recursive).

\smallskip

(ii) If the series $\sum_{l=1}^{\infty} R_l^{-1}$ converges and its sum is bounded by a constant $c$, then
$$
N_R(k,l)\le c(kR_l+1).
\eqno(1.2)
$$
\smallskip

(iii) If the series $\sum_{l=1}^{\infty} R_l^{-1}$ diverges, and  
$$
\sum_{l=1}^{M} R_l^{-1}\le F(M)
$$
for a certain increasing function $F=F_R$, then
$$
N_R(k,l)\le (kR_l+1) F(kR_l +1).
$$
}

\medskip

{\bf Proof.} The first statements are an easy exercise.
For the rest, notice that if $M$ is the minimal value for which
$(k,l)\in V_R(M)$, we have  $M-1<kR_l\le M$ and
$$
N_R(k,l) \le \r{card}\,V_R(M),
$$
and in the case (ii) we have 
$$
\r{card}\,V_R(M)\le \sum_{m=1}^{\infty} MR_m^{-1} \le c(kR_l+1).
$$

Similarly, in the case (iii) we have
$$
\r{card}\,V_R(M)\le M\sum_{m=1}^M R_m^{-1}\le (kR_l+1)F(kR_l+1)
$$
\smallskip

{\bf 1.6. L.~Levin's probability distributions.}  From (1.2) one sees that any sequence 
$\{R_l\}$ with
converging  $\sum_lR_l^{-1}$ can be used 
in order to construct the bijection
$\Z^+\times \Z^+\to \Z^+$, $(k,l)\mapsto N_R(k,l)$
linearly growing wrt $k$. Assume that it is computable
and therefore can play the role of $\theta$ in the proof of Theorem 1.4b).

\smallskip

In this case, 
 for any $l$, the set of rational numbers
$k/M\le r_l:=R_l^{-1}$ must be decidable. 

\smallskip

Even if we  weaken the last condition, requiring only enumerability
of the set $\{ (k,M,l) \,|\, k/M\le r_l\}$ (in particular, asking each $r_l$ to be computable from below), the convergence of $\sum_l r_l$
implies  that there is a universal upper bound (up to a constant) for such 
$r_l$. Namely, let $KP$ be  the exponentiated {\it prefix Kolmogorov complexity}
on $\Z$ defined with the help of a certain optimal prefix
enumeration (see [LiVi], [CaSt] for details).

\medskip

{\bf 1.6.1. Proposition.} ([Le]). {\it  For any sequence of
 numbers $r_l\ge 0$ with  enumerable set $\{ (k,M,l) \,|\, k/M\le r_l\}$ and convergent
$\sum_l r_l$, there exists a constant $c$ such that for all $l$,
$r_l\le c\cdot KP(l)^{-1}$ }

\medskip

More generally, L.~Levin constructs in this way a hierarchy
of complexity measures associated with  a class of abstract
norms, functionals on sequences computable from below.

\medskip

As I explained in the Introduction, this paper suggests that these {\it mathematical} distribution laws might 
lead to a new explanation of statistic properties of some  observable data. 

\bigskip

\centerline{\bf 2.~The computability (pro)perads and admissible families}

\medskip

{\bf 2.1. Libraries, oracles, and operators.} In this section, we will define 
admissible sets of partial $(m,n)$ functions $\Phi$ formalizing intuitive notions of
``software libraries and their reuse'' and ``oracle--assisted computation''.
A Kolmogorov complexity  relative to such a set will include
a formalization of the intuitive notion of relative complexity
$K(f|g)$ in the cases when $f,g$ are recursive (``reuse of $g$'')
and when $g$ might be uncomputable (``oracle--assisted computation'').

\smallskip

In this section our main objects are {\it not functions}
but objects of higher types:
\smallskip

(i)  {\it Programs for computing functions}, eventually even {\it names}
of oracles telling us values of uncomputable functions, and programs including 
these names.

\smallskip

(ii) {\it Operators,} that is programs {\it computing certain functions whose arguments
and values are  themselves programs.}

\smallskip

The main reason for such shift of attention is this. Already the set of  partial recursive $(m,n)$--functions for $m\ge 1$ 
{\it is not a constructive world}, as well as its extensions with which we deal
here. To the contrary, the set of  programs calculating recursive functions  in a chosen programming language,
such as Turing machines, or texts in a 
lambda--calculus, is constructive, but endowed with {\it uncomputable
equivalence relation:  ``two programs compute one and the same function''.}
We will call such worlds ``programming methods''. 

\smallskip

More precisely,
let $X,Y$ be two constructive worlds. A {\it programming method}
is a constructive world $P(X,Y)$ given together with a map
$P(X,Y)\to ParSet (X,Y)$, $p\mapsto \overline{p}$, where 
$ParSet$ is the category of sets with partial maps as morphisms.
We will systematically put bar over the name of a program to denote  the function  $\overline{p}:\,X\to Y$
which $p$ computes.

\smallskip

A good programming method must have additional coherence properties.
For brevity, consider only infinite constructive worlds, and assume that $P$ computes all recursive isomorphisms.
 We can then extend $P$ to any two infinite constructive worlds, and it is natural
 to require the existence of at least two additional  programs/operators
 $$
 Ev \in P(X\times P(X,Y),Y),\quad \overline{Ev}(x,p):= \overline{p}(x)\ \roman{for}\  x\in X,\ p\in P(X,Y),
 \eqno(2.1)
 $$
$$
Comp:\, P(P(X,Y)\times P(Y,Z))\to P(X,Z),\quad \overline{ \overline{Comp}(f,g)}= \overline{g}\circ \overline{f}.
\eqno(2.2)
$$
We will call such objects as $Ev$ and $Comp$ operators and say that they  {\it lift} the respective operations
on functions: evaluation at a point and composition.

\smallskip

For more detailed mathematical background, cf. [Man1], Ch.~IX, sec. 3--5.
\smallskip
In the following we start with a certain programming method $P$ computing
at least all partial recursive $(m,n)$--functions, and describe ways of extending it
necessary to  formalize the notions of library reuse and oracles.

\medskip
{\bf 2.2.~Constructing admissible sets.} Each such set $\Phi$ will be defined in the following way.
\smallskip

(i) Choose a constructive world $S$ consisting  of programs for computing  $(m,n)$--functions. 
It will be a union of two parts: (programs of) {\it elementary functions} and {\it library functions.}
All elementary functions will be (Turing) computable, i.~e.~(partial) recursive.
Library functions must form a constructive world (possibly, finite),
with some fixed numbering. The number of a library program is called its {\it address.}
Both library functiona and elementary functions will be decidable subsets of $S$.

\smallskip

(ii) Define a set of operators that can be performed on finite strings of
partial $(m_i,n_i)$ functions. They will be obtained by iterating several basic operators, such as 
$Comp$ and $Ev$ above. 
The world $OP$ of programs/names of such  operators will be
easy to encode by a certain constructive world of labelled graphs.

\smallskip

(iii) Take an element $P\in OP$ and specify a finite string $s:=(f_1,\dots ,f_r)$ of (addresses) of functions  from $S$
that can serve as input to $P$. The pair $(P, s)$ is then a program for calculation of a
concrete  string of $(m,n)$--functions.

\smallskip

(iv) Finally, $\Phi$ will be defined as the minimal set of functions computable by the programs in $S$
and closed with respect to the applications of all operators in $OP$.

\smallskip

We will now give our main examples of objects informally described in (i)--(iv).

\medskip

{\bf 2.3. The language of directed graphs.} In our world $OP$, each operator $\rho$  can take as input
a finite sequence of partial functions $f_i:\, (\Z^{+})^{m_i} \to  (\Z^{+})^{n_i}$, $m_i,n_i\ge 0$, $i=1,\dots, k)$
and produce from it another finite sequence of partial functions
$g_i:\, (\Z^{+})^{p_i} \to  (\Z^{+})^{q_i}$, $m_i,n_i\ge 0$, $i=1,\dots, l)$.
We will call {\it the signature} of $\rho$ the family
$$
sign\,(\rho ):= [(m_1,n_1),\dots ,(m_k,n_k); (p_1,q_1),\dots ,(p_l,q_l)] .
\eqno(2.3)
$$
Two operations $\rho ,\sigma$ can be composed if their signatures match:
$\rho\circ \sigma$ takes as input the output of $\sigma$.
\smallskip
Below we will describe explicitly a set of  {\it basic} operations $OP_0$. After that the whole set $OP$
will be defined as the minimal set of operations containing $OP_0$ and closed under composition.
\smallskip
A visually convenient representation of elements of $OP$ is given by (isomorphism classes of) directed
labeled graphs: see formal definitions in [BoMan], Section 1.
\smallskip
More precisely, each basic operator $\rho$ of signature (2.3) is represented
by a {\it corolla}:  graph with one vertex labelled by $\rho$; $k$ flags oriented towards the vertex (inputs) and 
$l$ flags oriented from the vertex (outputs) .  Moreover, inputs and outputs
must be totally ordered, and labelled by respective pairs $(m_i,n_i)$.
\smallskip

Labelled directed graphs with more vertices are obtained from a disjoint finite set of corollas
by grafting some outputs to some inputs. Grafted pair (output, input) must have equal labels $(m,n)$.
Flags that remain ungrafted form the inputs/outputs of the total graph.
\smallskip
Notice that directed graphs ([BoMan], 1.3.2) do not admit oriented loops.

\medskip

{\bf 2.4. Basic operators.}   In this subsection, we describe operations on strings of functions
that must be represented by the respective operators.  For brevity, we will denote by single
Greek letters these operators.

\smallskip

(a) {\it Composition of functions}.  This basic operator, say $\gamma$,  has signature of the form $[(m,n),(n,q); (m,q)]$.
It produces from two partial functions  $f:\, (\Z^{+})^{m} \to  (\Z^{+})^{n}$  and   $g:\, (\Z^{+})^{n} \to  (\Z^{+})^{q}$
their composition $g\circ f:\,  (\Z^{+})^{m} \to  (\Z^{+})^{q}$. Recall that $D(g\circ f):= f^{-1}(D(g))$.
It is a special case of operator $Comp$.

\medskip

(b) {\it Juxtaposition.}  This  basic operator, say $\sigma$,  has signature of the form
$$
[(m,n_1),\dots ,(m,n_k); (m, n_1+\dots +n_k)].
$$
It produces from $k$ partial functions  $f_i:\, (\Z^{+})^{m} \to  (\Z^{+})^{n_i}$  
the function $(f_1,\dots ,f_k)$: 
$$
D((f_1,\dots ,f_k))=D(f_1)\cap\dots \cap D(f_k),
$$
$$
(f_1,\dots ,f_k)(x_1,\dots ,x_m)= (f_1(x_1,\dots ,x_m),\dots , f_k(x_1,\dots ,x_m)).
$$

\smallskip

(c) {\it Recursion.} This  basic operator, say $\rho$,  has signature of the form
$$
[(m,1), (m+2,1); (m+1, 1)]
$$
It produces from  partial functions  $f:\, (\Z^{+})^{m} \to  \Z^{+}$  and   $g:\, (\Z^{+})^{m+2} \to  \Z^{+}$
the function $h:\, (\Z^{+})^{m+1} \to  \Z^{+}$such that
$$
h(x_1, \dots , x_m,1)= f(x_1,\dots , x_m),
$$
$$
h(x_1, \dots , x_m,k+1)= g(x_1,\dots , x_m, k, h(x_1,\dots ,x_m,k)))
$$
for $k\ge 1$.
\smallskip
The definition domain $D(h)$ is also defined by recursion:
$$
(x_1, \dots , x_m,1)\in D(h)  \Leftrightarrow  (x_1,\dots , x_m) \in D(f),
$$
$$
(x_1, \dots , x_m,k+1) \in D(h) 
$$
$$
\Leftrightarrow (x_1, \dots , x_m,k) \in D(h)\ \roman{and}\  (x_1,\dots , x_m, k, h(x_1,\dots ,x_m,k))\in D(g)
$$
for $k\ge 1$.

\medskip

(d) {\it Operator  $\mu$.} Its signature is $[(n+1,1); (n,1)]$.  Given an $(n+1,1)$--function
$f$,  it produces the $(n,1)$--function $h$ with the definition domain
$$
D(h)=\{(x_1,\dots ,x_n)\,|\, \exists x_{n+1}\ge 1\ \roman{such\ that}\
$$
$$
f(x_1,\dots ,x_n,x_{n+1})=1\ \roman{and}\ (x_1,\dots ,k)\in D(f)\ \roman{for\ all}\ k\le x_{n+1} \}.
$$
At the definition domain
$$
h(x_1,\dots ,x_n)=\roman{min}\,\{x_{n+1}\,|\, f(x_1,\dots ,x_{n+1})=1\}.
$$
\smallskip

(e) {\it Identity operation $\iota$.}
\medskip

{\bf 2.5. The constructive world of operations $OP$.} Returning now to sec. 2.2, 
we define $OP$ as the set of finite directed labelled graphs with totally
ordered inputs/outputs at each vertex  satisfying the following condition:
 each vertex is labelled  by one of the letters $\gamma, \sigma , \rho , \mu , \iota$
and its inputs/outputs are labelled by the respective components of the relevant signature..

\smallskip

We can choose any one of the standard ways to encode such graph by a string over a fixed finite alphabet,
and then define a structure numbering of $OP$ by ranking these words in alphabetic order.
The subset of  well--formed strings that encode graphs  ought  to
form a decidable subset of all strings, and all natural functions such as
$$
graph \mapsto sequence\ of\  all\ inputs\  of\ the\ graph\ with\ their\ (m,n)-labellings
$$
ought to be total recursive.

\smallskip

In the final count, each element of $OP$ determines an operation on  finite sets
of partial functions, producing another finite set of partial functions. Moreover, $OP$ can be enriched to
a free (pro)perad acting on finite strings of partial functions. If the signature of this string does not match
the signature of the inputs of the operation, we  may and will agree that  the operation produces an empty function.
The enrichment however requires some care and higher categorical constructions.

\medskip

{\bf 2.6. Basic partial functions.}  Let $S$ be a constructive world of programs/oracles
calculating partial $(m,n)$--functions. 
Denote by $OP(S)$ the minimal set of such programs
containing $S$ and closed with respect to application to them of operators from $OP$.
In the remainder of this paper, we will always include in $S$ a set of basic recursive functions $S_{rec}$,
such that $OP(S_{rec})$ consists of all (partial) recursive function.
In [Man1], Ch. V, Sec. 2,  the following set is chosen:
$$
\roman{suc}:\,\Z^+\to \Z^+,\quad x\mapsto x+1,
$$
$$
1^{(n)}:\,  (\Z^+)^m\to \Z^+, \quad (x_1, \dots x_m)\mapsto 1, \ n\ge 0.
$$
$$
\roman{pr}_i^m:\,  (\Z^+)^m\to \Z^+, \quad (x_1, \dots x_m)\mapsto x_i, \ n\ge 1.
$$

\smallskip

{\bf 2.7. Admissible sets and library reuse.} 
The standard Kolmogorov complexity of partial recursive functions is defined relative to the admissible set of functions
$\Phi$ computable by programs from $OP(S_{rec})$. If we include into the set $S$ only some programs
of recursive functions, then the total set of computable functions $\Phi$ will not grow,
but some functions will be computable by much shorter programs because the price of
writing a program for a library program can be disregarded.

\smallskip

{\bf 2.8. Admissible sets including oracles for uncomputable functions.} Here
one more complication arises: the requirement (ii) in our definition of the admissible sets
of functions is not satisfied automatically in the world $OP(S)$ as it was in the case
when the respective set of functions consisted only on recursive functions.
\smallskip

In order to remedy this, we have to add to the list of basic operators the operator $Ev$
from  (2.1). Its participation in the iteration of our former basic operations cannot be 
in an obvious way described by labelled graphs,
so more systematic  treatment is required: seemingly, we are in the realm of ``expanding
constructive universe'', some propaganda for which was made in [Man1], Ch. IX, sec. 3.

\bigskip

\centerline{\bf References}

\medskip

[BoMan] D.~Borisov, Yu.~Manin.  {\it Generalized operads and their inner cohomomorhisms.}
 In: Geometry and Dynamics of Groups
and spaces (In memory of Aleksader Reznikov). Ed. by M. Kapranov et al.
Progress in Math., vol. 265. 
Birkh\"auser, Boston, pp. 247--308.
Preprint math.CT/0609748

\smallskip
[CaSt] Ch.~S.~Calude, L.~Staiger. {\it  On universal computably enumerable prefix codes.}
 Math. Struct. in Comput. Sci. 19 (2009), no. 1, 45--57. 
\smallskip

[De]  S.~Dehaene. {\it The Number Sense. How the Mind creates Mathematics.}
Oxford UP, 1997.

\smallskip

[DeMe] S.~Dehaene, J.~Mehler. {\it Cross--linguistic regularities
in the frequency of number words.} Cognition, 43 (1992), 1--29.

\smallskip

[Del] J.--P.~Delahaye. {\it Les entiers ne naissent pas \'egaux.}
Pour la Science, no.~421, Nov.~2012, 80--85.

\smallskip

[FrChSh] K.~M.~Frahm, A.~D.~Chepelianskii, D.~L.~Shepeliansky.
{\it PageRank of integers.} arXiv:1205.6343
\smallskip

[HuYeYaHu] Shi-Ming Huang, David C.~Yen, Luen-Wei Yang, Jing-Shiuan Hua. {\it An investigation
of Zipf's Law for fraud detection. } Decision Support Systems, 46 (2008), 70--83.

\smallskip
[Le] L.~Levin. {\it Various measures of complexity for finite objects
(axiomatic description).} Soviet Math. Dokl., vol 17, No. 2 (1976), 522 --526.
\smallskip

[LiVi] Ming Li, P.~Vit\'anyi. {\it An introduction to Kolmogorov complexity
and its applications.} Springer, 1993.

\smallskip

[Mand]  B.~Mandelbrot. {\it An information theory of the statistical
structure of languages.} In Communication Theory (ed. by W.~Jackson, 
pp. 486--502, Butterworth, Woburn, MA, 1953.

\smallskip
[Ma1] D.~Yu.~Manin. {\it Zipf's Law and Avoidance of Excessive Synonymy.} Cognitive
Science, vol.~32, issue 7 (2008), pp. 1075--1078. arXiv:0710.0105.

\smallskip

[Ma2]  D.~Yu.~Manin. {\it Mandelbrot's model for Zipf's Law. Can  Mandelbrot's model explain Zipf's Law
for language?} Journ. of Quantitative Linguistics, vol.16, No. 3 (2009), 274--285.

\smallskip

[Man1] Yu.~I.~Manin. {\it A Course in Mathematical Logic for Mathematicians.} Second Edition.
Graduate  Texts in Mathematics, Springer Verlag, 2010.

\smallskip

[Man2] Yu.~Manin. {\it Renormalization and computation II. } Math.~Struct.~in Comp.~Sci., vol. 2,
 pp.~729--751, 2012, Cambridge UP.
 arXiv:math.QA/0908.3430
\smallskip

[Man3] Yu.~Manin. {\it Kolmogorov complexity
as a hidden factor of  scientific discourse:
from Newton's law to data mining.} 
Talk at the Plenary Session of the Pontifical Academy of Sciences on
``Complexity and Analogy in Science: Theoretical, Methodological and Epistemological Aspects'' , Vatican, 
November 5--7, 2012. arXiv:1301.0081

\smallskip

[ManMar] Yu.~Manin, M.~Marcolli. {\it Kolmogorov complexity and the asymptotic bound for error-correcting
codes}.  arXiv:1203.0653

\smallskip

[MurSo] B.~C.~Murtra, R.~Sol\'e. {\it On the Universality of Zipf's Law.}  (2010),
Santa Fe Institute.{\it  (available online).}

\smallskip

[NaWe] A. Nabutovsky, S.~Weinberger. {\it The fractal nature of
Riemm/Diff I.} Geometriae Dedicata, 101 (2003), 145--250.
\smallskip

[Ro] H.~Rogers. {\it G\"odel numberings of partial recursive functions.}
Journ. Symb. Logic, 23 (1958), 331--341.

\smallskip

[Sch] C.~P.~Schnorr. {\it Optimal enumerations and optimal G\"odel
numberings.} Math. Systems Theory, vol. 8, No. 2 (1974), 182--191.

\smallskip
[Ta] T.~Tao. {\it E pluribus unum: From Complexity, Universality.} Daedalus, Journ. of the AAAS, Summer 2012, 
23--34.

\smallskip

[Ve] Todd L.~Veldhuizen. {\it Software Libraries and Their Reuse: Entropy, Kolmogorov Complexity,
and Zipf's Law.} arXiv:cs/0508023

\smallskip

[Ya] N.~S.~Yanofsky. {\it Towards a definition of an algorithm.}   J.~Logic Comput. ~21 (2011), no. 2, 253--286.
math.LO/0602053

\smallskip

[Zi1] G.~K.~Zipf. {\it The psycho--biology of language.} London, Routledge, 1936.

\smallskip
[Zi2] G.~K.~Zipf. {\it Human behavior and the principle  of least effort.} Addison--Wesley, 1949.
\smallskip

[ZvLe]  A.~K.~ Zvonkin,  L.~A.~ Levin. {\it
The complexity of finite objects and the basing of the concepts of information and randomness on the theory of algorithms. }(Russian)
Uspehi Mat. Nauk 25,  no. 6(156) (1970), 8--127.

\enddocument